\documentclass{article}
\usepackage[dvips,final]{graphicx}
\usepackage{amsmath}  %use package for writing math eqms
\usepackage{graphicx} %use package for inserting images
\usepackage{fancybox} %use package to help with creating boxed equations
\usepackage{mathrsfs} %use package to insert script letters in math eqns
\usepackage{amssymb} %use for math symbols
\usepackage{setspace}
\usepackage{verbatim}
\usepackage{authblk}
\newcommand{\eqn}[1]{\begin{equation}#1\end{equation}}

\newcommand{\e}[1]{e^{#1}}
\newcommand{\cyvol}{\mathcal{V}}

\newcommand{\mplanck}{M_P}
\newcommand{\lagrange}{\mathcal{L}}

\newcommand{\kahler}{K\"ahler }
\newcommand{\etapar}{\eta^{\parallel}}
\newcommand{\etaperp}{\eta^{\perp}}
\newcommand{\window}{\mathcal{W}}
\newcommand{\ud}{\,\mathrm{d}}
\newcommand{\xiperp}{\xi^{\perp}}
\newcommand{\xipar}{\xi^{\parallel}}

\newcommand{\fNL}{f_{NL}}

\begin{document}

\title{Curvature Spectra and Nongaussianities in the Roulette 
Inflation Model}
\author[1]{Aaron C. Vincent\footnote{vincenta@hep.physics.mcgill.ca}}
\author[1]{James M. Cline\footnote{jcline@hep.physics.mcgill.ca}}
\affil[1]{\textit{Department of Physics, Rutherford Physics Building, McGill University, 3600 Rue University, Montreal, QC, Canada H3A 2T8.}}
\maketitle

% \begin{center}
%  \emph{Department of Physics, Rutherford Physics Building, McGill University, 3600 Rue University, Montreal, QC, Canada H3A 2T8.}
% \end{center}
% 
% 

\bibliographystyle{unsrt}

%%%%%%%%%%%%%%%%%%%%%%%%%%%%%%%%%%%%%%%%%%%%%%%%%%%%%
%%         English Abstract                        %%
%%%%%%%%%%%%%%%%%%%%%%%%%%%%%%%%%%%%%%%%%%%%%%%%%%%%%

\begin{abstract} Using the gradient expansion method of Rigopoulos,
Shellard and van Tent which treats cosmological perturbations as gradients on top of a
homogeneous and isotropic FRW background, we study the production of
nongaussianities in the roulette model of inflation. % \cite{roulette}.
Investigating a number of trajectories within this two-field model of
inflation, we find that while the superhorizon influence of the
isocurvature modes on the curvature bispectrum produces nonzero
contribution to $\fNL$, the effect is negligible next to the standard
inflationary prediction $|\fNL| \sim n_s - 1$. % \cite{maldacena}. 
This
is the case in both the squeezed and equilateral configurations of
the bispectrum, although the former is slightly larger in the
trajectories under consideration.  \end{abstract}

\section{Introduction}

Recent advances in vacuum stabilization in string theory, most
notably the KKLT \cite{KKLT} and the large volume  \cite{largevol}
varieties, have put stringy realizations of cosmic inflation on a much
firmer footing.  Only when all moduli have been stabilized does it
make sense to study which of them might be candidates for inflation.
String theory provides an abundance of moduli fields, for example the
size and shape of the extra dimensions, and thus many potential
opportunities for inflation.  These include single-field models, such
as ref.\ \cite{kahler}, and multiple-field models  such as the
racetrack model \cite{racetrack}, the N-flation scenario
\cite{nflation}, or the swiss-cheese scenario \cite{swiss}.  Multiple field inflation has the advantage of  more
easily providing some of the nonstandard observational features that
one would like for helping to discriminate between theories. The
extra  degrees of freedom allow the production of isocurvature modes
(perturbations transverse to the classical trajectory) and it has become
apparent \cite{largenongauss} that these may give rise to large
nongaussianities in the cosmic microwave background (CMB) temperature
fluctuations.

One such model is the roulette scenario \cite{roulette}, in which the
Calabi-Yau manifold (the compactified extra dimensions of string
theory) relaxes from an initial excited state towards a  minimum of
its potential. The large volume compactification that is used ensures
that this minimum exists for large ranges of the microscopic
parameters. In addition, unlike KKLT, it does not require  tuning the
constant term in the superpotential to very small values, and it
gives a natural expansion parameter, the inverse volume $1/\cyvol$,
providing a controlled $\alpha'$ expansion. In this particular model, the last four-cycle and
 corresponding axionic partner to relax act as slow rolling scalar fields which
drive the final stage of inflation. 

Specializing to a specific set of microscopic parameters, we explored the various trajectories of inflation available within the context of this two-field model. The prediction of observable parameter values were in accordance with known results from
CMB and large scale structure survey data. The influence of isocurvature perturbations, which seed inhomogeneieties between the species of fields driving inflation, was furthermore found to be quite important. 

In addition, recent claims of detection of nongaussianity in the WMAP
CMB data, specifically a nonvanishing nonlinearity parameter $f_{NL}$
\cite{yadav} have sparked a large interest in deviations from
gaussianity in the spectrum of primordial fluctuations as an
additional observable that must be predicted by a successful theory
of the early universe. Given that precision measurements of $f_{NL}$
from experiments such as Planck will soon be available, it is all the
more important that the mechanisms governing the production and
evolution of primordial nongaussianities be well understood.

We will first provide an overview of the roulette model, followed by a discussion of primordial nongaussianities from inflation. In Section \ref{sec:results} we will give the main results of our paper, which are the numerical caluclations of nongaussianities in the curvature perturbation. The important aspects of our calculations, which follow \cite{quantitative}, are presented in Appendix A.

\section{Roulette (K\"ahler Moduli) inflation}
\label{sec:roulette}
	
\begin{figure}[ht]
\centerline{\includegraphics[width=0.9\textwidth]{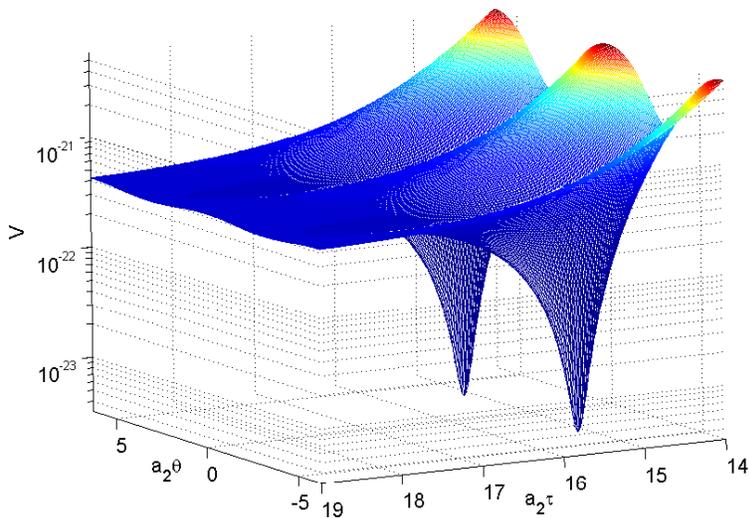}}
\caption[roulette inflation potential]{The potential \eqref{effpot} as of function of the volume 
modulus $\tau$ and its axionic partner $\theta$ for parameters
in Table 2. The potential is periodic in the $\theta$ direction.} 
\label{potential1}
\end{figure}

The roulette model is a string theoretic inflationary scenario set in
the context of a Type IIB large volume compactification. Although
there may be evolution of several K\"ahler moduli, the observable
part of inflation is governed by last (and lightest) one to relax.
Since the earlier-evolving moduli stabilize to deep minuma, they
rapidly decouple from the dynamics \cite{roulette}, before the final
60 $e$-foldings. The name ``roulette'' comes from the cyclic shape of
the potential, resembling a roulette table whose grooves are the
minima toward which the inflaton eventually relaxes (Figure \
\ref{potential1}). During
inflation, the F-term potential of the large volume compactification
is flat enough to allow slow-rolling over sizeable patches of field
space. Reheating, which we will not address here, occurs when the
inflaton fields oscillate at the bottom of the potential. This model was first proposed as a single-field inflation model by Conlon and Quevedo in \cite{kahler}, and subsequently generalized to include the axion as a second inflaton field by Bond, Kofman, Prokushkin and Vaudrevange in ref.\ \cite{roulette}. As in ref.\ \cite{roulette}, we use the large-volume compactification
\cite{largevol}, in which the 10 spacetime dimensions of type IIB string theory are separated into a 4-dimensional noncompact spacetime and a conformally Calabi-Yau 3-fold.

After minimizing the F-term potential in the large volume compactification with respect to the axio-dilaton and all \kahler moduli but $T_2 = \tau_2+i\theta_2$, the potential can be expressed as a function of  $\tau \equiv \tau_2$
and $\theta \equiv \theta_2$. After expanding in powers of $1/\cyvol$
this potential reduces to \cite{roulette}:
\eqn{
	V = \frac{8(a_2A_2)^2\sqrt{\tau}e^{-2a_2\tau}}
{3\alpha\lambda_2 \cyvol} + \frac{4W_0a_2A_2\tau e^{-a_2 \tau}
 \cos(a_2\theta)}{\cyvol^2} +\Delta V + O(1/\cyvol^3),
\label{effpot}
}
where $\Delta V$ is the uplifting contribution, of order $1/\cyvol^2$
\cite{kahler}, adjusted so that 
that $V=0$ at its minimum.   It is crucial for the naturalness of K\"ahler
moduli inflation that $\Delta V$ depends only very weakly upon
$\tau_2$ in the case of interest, where $\tau_1\gg\tau_2$.  Because
of this, $\Delta V$ is nearly constant during inflation, and the
slow-roll condition on $\tau_2$ is easily satisfied.
Fig.\ \ref{potential1} shows the form
 of the potential.

For notational purposes we define $\phi^1 = \tau \equiv \tau_2$ and
$\phi^2= \theta \equiv \theta_2$. The kinetic term takes the diagonal form:
\eqn{
	\lagrange_{\rm kin} = \frac12 K_{2\bar{2}}\delta_{AB}
\partial_{\mu}\phi^A \partial^{\mu}\phi^B,
}
with the $(2,\bar{2})$ component of the \kahler metric given by \cite{roulette}
\eqn{
K_{2\bar{2}} = \frac{3 \alpha \lambda_2[2\cyvol + \xi + 6\alpha
 \lambda_2 \tau_2^{3/2}]}{4(2\cyvol+\xi)^2\sqrt{\tau_2}}.
\label{k22}
}

In the original \kahler moduli inflation model \cite{kahler} the Standard Model was confined to a D7-brane that wraps the inflaton cycle. It has recently been pointed out, however, \cite{fibre} that string loop corrections contribute to the \kahler potential in such a way that the exponential flatness of $V$ is destroyed. This would give large contributions to the $\eta$ parameter, thereby preventing slow roll from occuring. Ref. \cite{fibre} does point out that this can be remedied by removing the D7 brane from the inflating cycle, which has the disadvantage of complicating the reheating process. This could proceed via a mechanism akin to the ones studied in multiple-throat inflationary models in warped compactifications. See for instance ref.\ \cite{mthroat}. Given that we are only concerned with the inflationary phase in the present Paper, we will take the potential \eqref{effpot} as is.

\section{Inflation and nongaussianities}
\label{sec:nongauss}

Slow-roll inflationary scenarios generically predict a near-scale
invariant spectrum of primordial perturbations, with nearly Gaussian
statistics. This is expected, since individual quantum fluctuations can be treated as independent results and should therefore be Gaussian by virtue of the central limit theorem. 
However, more complex field interactions, along with interactions with gravity, which is inherently non-linear, are sure to produce at least small deviations from Gaussianity. Due to the stochastic nature of these perturbations, it is therefore natural to develop statistical tools to compare predictions of the theory with CMB observations. Wick's theorem tells us that the even moments ($2n$-point correlators) of a linear field or distribution $\phi_L$ decompose into a sum over the permutations of two-point correlators, whereas the odd moments vanish. The measurement of deviations from gaussianity in a field $\phi(\textbf{x})$, can therefore be made through the bispectrum (the Fourier transform of the three-point correlator), and through the connected part of the trispectrum, that is, the part of the Fourier transform of the four-point correlator that cannot be decomposed into products of the power spectrum. 

 It is common to parameterize the small deviations from 
gaussianity in terms of their effect on the Bardeen potential
$\Phi$ through $f_{NL}$ \cite{acoustic}
\begin{equation}
 \Phi(\mathbf{x}) = \Phi_L(\mathbf{x})  + 
f_{NL}\left(\Phi_L^2(\mathbf{x}) - \langle \Phi_L^2(\mathbf{x}) 
\rangle \right),
\label{localfnl}
\end{equation}
where $\Phi_L$ is a purely gaussian random field with
$\langle\Phi_L\rangle = 0$. Even for large $f_{NL}$ this parameterisation is sufficient, given that the fluctuations $\Phi$ are order $\sim 10^{-5}$. Already with the COBE observations it
was shown that the nongaussian fraction of $\Phi$ must be less than a
few percent, $f_{NL}\langle\Phi_L^2\rangle^{1/2} < 0.04$ (see for
example \cite{Komatsu:2003fd}).  Subsequent measurements have
tightened this limit to the level of 
$f_{NL}\langle\Phi_L^2\rangle^{1/2} < 0.003$.  Some recent published
limits on $f_{NL}$ are shown in Table \ref{cmbfnl}.

\begin{table}
\caption{Some recent $95\%$ CL estimates of $f_{NL}$ using WMAP data.}
\begin{center}
\begin{tabular}{|c l l| c |}
\hline
Komatsu \textit{et al.} & (WMAP 1-year) &\cite{Komatsu:2003fd} & $-58 < f_{NL} < 134$ \\ \hline 
Creminelli \textit{et al.} & (WMAP 3-year)&\cite{creminelli}	&	$-36 < f_{NL} < 100$ \\ \hline
Yadav and Wandelt & (WMAP 3-year)&\cite{yadav}		&$27 < f_{NL} < 147$ \\ \hline
Komatsu \textit{et al.} & (WMAP 5-year) &\cite{wmap5}	& $-9 < f_{NL} < 111$ \\ \hline
\end{tabular}
\end{center}
\label{cmbfnl}
\end{table}

Some of these results \cite{yadav} suggest that the CMB anisotropies
exhibit measurable deviations from Gaussian statistics. Whether or
not these detections are confirmed, future observations such as the
9-year WMAP data and the Planck satellite data will provide stringent
bounds on the primordial bispectrum, yielding  additional parameters that any
successful model of the early universe will have to match. Although
nongaussianities may be measured from the contribution of any $n > 2$
connected $n$-point function, the 3-point correlator is the easiest
to detect due to the smallness of the anisotropies.

In the original (``local'') ansatz (\ref{localfnl}), $f_{NL}$ was
taken to be a number, but by relating it to the bispectrum one sees
that more generally it could be a function of the momenta $k_i$.
Taking the Fourier transform and writing $\Phi(k) = \Phi_L(k) +
\Phi_{NL}(k)$, is  easy to show (noting that $\langle \Phi_L^3
\rangle$ vanishes identically) that the lowest order nonvanishing
component of $f_{NL}$ may be written in terms of the bispectrum and
power spectrum:\footnote{One must be careful with the
sign of $f_{NL}$, which has been a source of some confusion in the
literature, due to the sign difference between the Bardeen potential
and the gravitational potential 
(see appendix A2 of ref.\ \cite{LoVerde}.)
We use the WMAP convention that positive $f_{NL}$ corresponds to positive
bispectrum of $\Phi$.}
\eqn{
f_{NL}\ \sim \ \delta(\textbf k_1 + \textbf k_2 + \textbf k_3)\frac{\langle \Phi_L(k_1) \Phi_L(k_2) \Phi_{NL}(k_3)\rangle}{\langle \Phi_L(k_1) \Phi_L(k_2) \rangle^2}.
\label{fnldef}
}

Due to the delta function, the wave vectors form a triangle,
and the $k$-dependence of eq.\ (\ref{fnldef})
can be expressed in terms of  two ratios of momenta, for example
$k_2/k_1$ and $k_3/k_1$, and an overall scale. Different mechanisms
produce bispectra that peak for differently shaped triangles;
for example equilateral ($k_1 = k_2 = k_3$), 
or squeezed ($k_1 \ll k_2
= k_3$).   The latter corresponds to the prediction of local ansatz.
Single-field slow-roll inflation predicts nongaussianity
of the local type \cite{maldacena}, given that the dominant contribution to the bispectrum should come from the superhorizon influence of small $k$ modes which act to ``rescale'' modes as they evolve toward the end of inflation. A rigorous expansion of the action to third order in perturbation theory is given in ref.\ \cite{maldacena}. Other models such as ghost inflation and DBI inflation predict large $f_{NL}$ for the equilateral configuration \cite{shape}, in which non-gaussianities are created before horizon-crossing. 

Our focus will be on tracking the perturbative curvature modes from
the time they expand beyond the Hubble radius $H^{-1}$, until the end
of inflation. We will use the gradient expansion approach developed
by Rigopoulos, Shellard and van Tent
\cite{nonlin,largenongauss,quantitative}. The assumptions of
homogeneity and isotropy of the background FRW inflationary universe 
allow the use of the ``long-wavelength'' approximation, in which the 
gradient terms of the equations of motion may be dropped in the
classical (unperturbed) equations of motion \cite{nonlin,salopek}:
\begin{eqnarray}
	D_t\Pi^A + 3NH\Pi^A &=& -NG^{AB}V_{,B},
 \label{eom2}\\
 \partial_tH &=& -\frac{1}{2}N\,\Pi_A\Pi^A,
\label{eom3} 
\end{eqnarray}
where the Hubble rate is
\eqn{ H^2 =
\frac13\left(\lagrange_{\rm kin} + V\right),
\label{eom1}
}
on scales larger than the Hubble length $H^{-1}$. As a consequence,
the inclusion of metric and field perturbations simply amounts to the
inclusion of gradient terms on top of the background, whose equations
of motion are computed from the full field equations of motion. The
advantage of this approach is twofold: no slow-roll approximation is
needed to find and solve the equations of motion, and the resulting
equations are exact (non-perturbative) results. Quantitative
computation of power spectra and bispectra was done via a
perturbative expansion of these gradient equations of motion. The
important results of \cite{quantitative} that we made use of are
presented in Appendix A.  Here we will only sketch the method and refer the
reader interested in the details to the Appendix.

For the two-field case such as we consider here, the dynamical
degrees of freedom for the metric fluctuations can be reduced to
$(\zeta^1,\zeta^2,\theta^2)$ where $\zeta^1$ is the adiabatic
curvature fluctuation, $\zeta^2$ is the isocurvature fluctuation, and
$\theta^2 = \dot\zeta_2$.  ($\theta^2$ is like the canonical momentum
conjugate to $\zeta^2$, and the evolution equations are first order
in derivatives, similar to Hamilton's equations of motion.  $\theta^1$ is
not an independent degree of freedom, but is constrained in terms of
the others.)\ The formalism solves for the gradient  of
the three fields, $v_{ia} = \partial_i (\zeta^1,\zeta^2,\theta^2)$,
where $a=1,2,3$.  
These quantities can furthermore be expanded
order-by-order in the cosmological perturbation, $v_{ia} =
v_{ia}^{(1)} + v_{ia}^{(2)} + \dots$.  Higher order terms are sourced
by the next lowest order ones, through a master equation of the form
\eqn{
\dot{v}_{ia}(t,\textbf{x}) + A_{ab}(t,\textbf{x})v_{ib}(t,\textbf{x}) = 0.
\label{nl1}
} 
where the matrix $A_{ab}$ (see (\ref{Aeq}) is determined by various
slow roll parameters.  The lowest order source term from which higher
order fluctuations follow is deduced by the method of stochastic
quantization. This information is encoded in a matrix $X^{(1)}_{bm}$,
eq.\ (\ref{Xeq}).   The solutions of eq.\ (\ref{nl1}) for $v_{ia}^{(1)}$ 
and $v_{ia}^{(2)}$ which we will need for computing the spectrum and
bispectrum can be expressed in terms of a Green's function
$G_{ab}(t-t')$ which is the solution to 
\eqn{
\frac{\ud}{\ud t}G_{ab}(t,t') + A_{ac}^{(0)}(t)G_{cb}(t,t') = \delta(t-t').
\label{greens2}
}
where $A_{ac}^{(0)}$ is the matrix $A_{ac}$ evaluated using just the
homogeneous background solution.  The main technical difficulty then
is in computing the Green's function.  The fluctuation $v_{ia}^{(1)}$ 
can then be computed through
$v_{am}^{(1)}(k,t)  G_{ab}(t,t_* + \ln c)\,X^{(1)}_{bm}(k,t_* + \ln
c)$ (see eq.\ (\ref{veq})) and similarly $v_{ia}^{(2)}$ is given by
eq.\ (\ref{veq2}). The nongaussianity parameter $f_{NL}$ is determined
by $v_{ia}^{(1)}$ and $v_{ia}^{(2)}$ through eqs.\
(\ref{bispectrum}) and (\ref{fkkp}) of the appendix.

For the roulette model, we must solve these equations numerically. Analytic
results have been developed ({\it e.g.}, in ref.\  \cite{largenongauss}) in
the context of this formalism, but only within certain constrained
limits.

\section{Numerical method and results}
\label{sec:results}

In this section we will consider a number of different parameter sets
for the model, indicated in Table 2.  Special attention will be given
to the first of these, for which we illustrate the different
possibilities depending on which  inflationary trajectory is followed
({\it i.e.,} the dependence on  the initial conditions).  In each
case, we evolved the light fields $(\tau,\theta)$ starting from rest,
until the end of inflation, which we took to be the point at which
the slow roll parameter $\epsilon$ grew to $\epsilon=1$.  For each
trajectory, we solved for the perturbation amplitudes
${v}_{ia}^{(m)}$  to first and second order using the gradient method
described above. The power spectrum, scalar spectral index and
nonlinearity parameter $\fNL$ thus calculated are presented below for
a variety of inflationary trajectories and parameter choices. We also
discuss the superhorizon influence of the isocurvature modes on these
quantities in what follows. 

\begin{table}[b]
\caption[roulette parameter set]
{Parameter sets used for numerical simulations of the roulette model.
The first corresponds to parameter set 1 of \cite{roulette}, but with
$\cyvol$ adjusted to meet COBE normalization and was used in all
simulations unless otherwise indicated.}
\begin{center}
% use packages: array
\begin{tabular}{|c|c|c|c|c|c|c|c|}
\hline
Set & $W_0$ & $a_2$ & $A_2$ & $\lambda_2$ & $\alpha$ & $\xi$ & $\cyvol$  \\ \hline
1 & 300 & $2\pi/3$ & 0.1 & 1 & $1/9\sqrt 2$ & 0.5 & $8 \times 10^8$ \\ \hline 
2 & $6 \times 10^4$ & $2\pi/30$ & 0.1 & 1 & $1/9\sqrt 2$ & 0.5 & $10^8$ \\ \hline 
3 & $4 \times 10^5$& $\pi/100$ & 0.1 & 1 & $1/9\sqrt 2$ & 0.5 & $10^9$ \\ \hline 
4 & 200 & $\pi$ & 0.1 & 1 & $1/9\sqrt 2$ & 0.5 & $10^6$\\ \hline
\end{tabular}
\label{paramset}
\end{center}
\end{table}

\begin{figure}[h]
\begin{center}
 	\includegraphics[width = 0.85\textwidth]{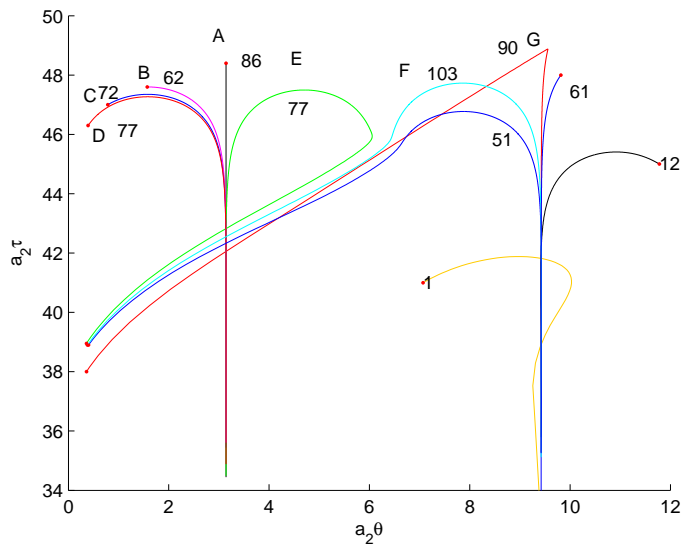}
\end{center}
\vskip-0.5cm
\caption[roulette inflation trajectories]{Various field trajectories
for different initial field configurations (red dots). The numbers
beside each curve are the number of e-foldings before slow roll
breaks down and $\epsilon$ exceedes 1. The potential here used the
parameters from Table \ref{paramset}. The labeled trajectories A
through G correspond to those listed in Table \ref{fNLtable}.} 
\label{trajectories}
\end{figure}

\begin{figure}[h]
	\includegraphics[width = 0.9\textwidth]{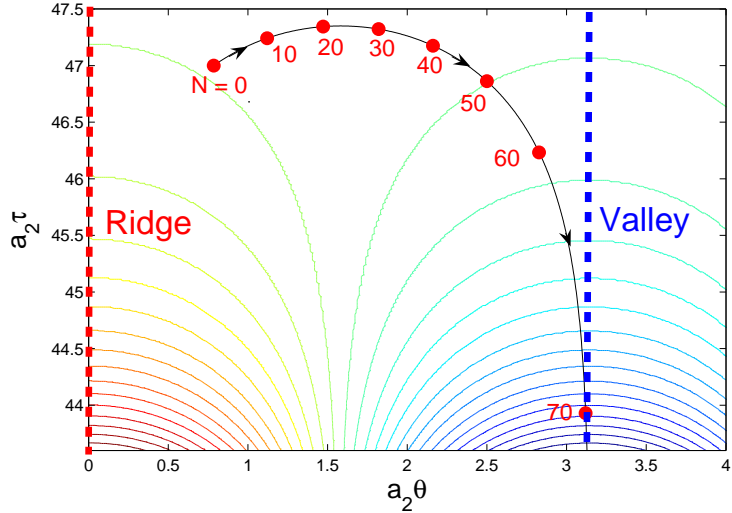}
\caption[Detailed roulette trajectory]{Detailed plot of trajectory C
from Figure \ref{trajectories}, superimposed on a contour plot of the
inflaton potential. Red dots represent time increments in e-foldings.} 
\label{contourTrajectory}
\end{figure}

In contrast with the KKLT compactifiation, the large volume scenario
places no strong restrictions on the value of $W_0$ in the effective
field theory \cite{largevol}. One therefore has considerable freedom
to vary the parameters of the potential \eqref{effpot}.  We first
verified the results of ref.\ \cite{roulette} as a check on
methodology. We solved for the inflation trajectories starting from a
variety of initial conditions and verified that slow roll  was
generically obtained, as found in section 6 of \cite{roulette}. Some
examples are illustrated in Figure \ref{trajectories}. Solution A,
the ``$\tau$-valley'' trajectory of Conlon and Quevedo \cite{kahler},
effectively corresponds to single-field inflation, as the fields
start with $\theta$ already minimized. Figure \ref{contourTrajectory}
shows a more detailed plot of one of the trajectories, superimposed
on a contour plot of the potential. 
Values of $\epsilon$ were consistently very small, with $\log\epsilon
\sim -13$ at the COBE scale. Typical values of the tensor-to-scalar
ratio produced by the fields were therefore $r \simeq 3.5 \times
10^{-12}$.

We focused attention on  parameter set 1 of \cite{roulette}, with the
modification that the volume $\cyvol$ was tuned to achieve COBE
normalisation ($\cyvol = 8 \times 10^8 l_s^6$; see Table
\ref{paramset}), {\it i.e.,} the power spectrum is $P_s \sim 4 \times
10^{-10}$ on COBE scales, in order to have a realistic example. This
corresponds to an inflationary energy scale  $V^{1/4} \simeq 10^{13}$
GeV, giving a duration of $52$-$55$ observable e-foldings of
inflation, assuming reheating temperatures of $10^{10}$-$10^{13}$ GeV. Although a more generic method of normalization is to
rescale the potential by an overall factor, the dependence on
$1/\cyvol$ of both terms in the potential gives a way to adjust its
magnitude without introducing additional parameters.

\begin{table}[h]
\caption{Cosmological observables computed from chosen trajectories
(see figure \ref{trajectories}). The top part uses parameter set 1,
whereas the bottom part displays results from the other sets (see
Table \ref{paramset}). $f_{NL}^{\rm equil.}$ and $f_{NL}^{\rm sqzd.}$ are the
nonlinearity parameter in the equilateral and squeezed
configurations, respectively; $N$ is the total number of e-foldings
of inflation;  $n_s^{\rm s.f.} = 2\eta - 6\epsilon$ is the standard
single-field scalar spectral index, whereas $n_s^{\rm exact}$ is the
exact index. $p^{\mathrm{iso}}$ is the proportion of the curvature
power spectrum originating from the isocurvature modes during
superhorizon evolution, calculated using \eqref{piso}. All
observables are computed for the mode $k$ that left the horizon
55 e-foldings before the end of inflation, and are evaluated at the end
of inflation.}
\begin{center}
\begin{tabular}{|c|c|c|c|c|c|c|c|c|}
\hline
Trajectory &$\mathbf{a_2\tau}$ & $\mathbf{a_2\theta}$ &
$\mathbf{f_{NL}^{equil.}}$ & $\mathbf{f_{NL}^{sqzd.}}$ &
$\mathbf{N}$ & $\mathbf{n_s^{s.f.}} $ & $\mathbf{n_s^{exact}}$ & $\mathbf{p^{\mathrm{iso}}}$ \\ \hline
A &48.4 & $\pi$ & 0 & 0 & 86 & 0.965 & 0.965 & 0\\ \hline
B &47.6 & $\pi/2$ & 0.0052 & 0.0069 & 62 & 0.992 & 0.976 & 0.70 \\ \hline
C &47.0 & $\pi/4$ & 0.0105 & 0.0132 & 72 & 1.02 & 0.940 & 0.81\\ \hline
D &46.3 & $\pi/8$ & 0.0242 & 0.0272 & 77 & 1.02 & 0.920 & 0.91\\ \hline
E &38.95 & 0.36 & 0.0688 & 0.0698 & 77 & 1.04 & 0.930 & 0.90 \\ \hline
F &38.93 & 0.40 & 0.0068 & 0.0089 & 103 & 1.002 & 0.948 & 0.72\\ \hline
G &38 & 0.38 & -0.0060 & -0.0060 & 90 & 0.964 & 0.965 & 0.02\\ \hline
\hline
Param.\ set &$\mathbf{a_2\tau}$ & $\mathbf{a_2\theta}$ &
$\mathbf{f_{NL}^{equil.}}$ & $\mathbf{f_{NL}^{sqzd.}}$ & $\mathbf{N}$
& $\mathbf{n_s^{s.f.}} $ & $\mathbf{n_s^{exact}}$ & $\mathbf{p^{\mathrm{iso}}}$ \\ \hline
2 &29.5 & $\pi/32$ & 0.0301 & 0.0338 & 77 & 1.055 & 0.909 & 0.93 \\ \hline
3 &28.7 & $\pi/32$ & 0.0553 & 0.0590 & 69 & 1.1045 & 0.849 & 0.97 \\ \hline
4 &33.2 & $\pi/32$ & 0.0404 & 0.0442 & 73 & 1.074 & 0.891 & 0.93 \\ \hline
4 &35.0 & $\pi/8$ & 0.0060 & 0.0082 & 94 & 1.002 & 0.946 & 0.73 \\ \hline

\end{tabular}
\end{center}
\label{fNLtable}
\end{table}
 
\subsection{Isocurvature Perturbations}

 In the case of more highly curved trajectories, the isocurvature
modes were found to have a large, positive effect on the power
spectrum of adiabatic perturbations, and consequently on the scalar
spectral index $n_s$, as recently shown in \cite{lalak}. Large
curvature in field space during the course of inflation resulted in a
``projection'' of the isocurvature modes onto the adiabatic
direction. For some trajectories with long periods of curving,  over
90\% of the power spectrum originated from the isocurvature modes.
Table \ref{fNLtable} gives the proportion of the observable curvature
power spectrum at the end of inflation that results from the
influence of the isocurvature mode:
\eqn{
	p^{\mathrm{iso.}} \equiv \left| \frac{P_{\mathrm{{\rm
s.f.}}}-P_{\mathrm{exact}}}{P_{\mathrm{exact}}} \right|_{t_* = 55},
\label{piso}
}
where the subscript $t_* = 55$ indicates that these quantities were
evaluated  at the COBE scale, which we take to be the modes that
crossed the horizon 55 e-foldings before the end of inflation.
$P_{\mathrm{{\rm s.f.}}}$ is the power spectrum computed with the
effective single-field result
\eqn{
	P_{\mathrm{{\rm s.f.}}}  = \frac{1}{50 \pi^2} \frac{H^4}{\lagrange_{\mathrm{kin}}},
}
whereas $P_{\mathrm{exact}}$ was computed using \eqref{fullP}. A
second result of the influence of isocurvature modes (also discussed
in ref.\ \cite{lalak}) is a lower scalar spectral index than would be
na\"{\i}vely expected from the single-field result $n_s^{\rm s.f.} = 2\eta -
6\epsilon$ ($ = -4\epsilon - 6\etapar$ in the notation of Appendix
A). The full scalar spectral index $n_s$ at COBE scales was computed
by taking the derivative of a cubic fit of $\ln{P_s}$, with
the power spectrum evaluated from \eqref{fullP}:
\eqn{
n_s^{(\rm exact)} = \frac{d \ln{P_{\mathrm{exact}}}}{d\ln k} = \left.
\frac{d \ln{P_{\mathrm{exact}}}}{dt} \right\vert_{t_* = 55}.
}
Here we have used the fact that in the gauge $NH = 1$ with 
$H$ approximately constant, 
${d}/{d\ln k} = {d}/{d \ln(aH)_*} \simeq H^{-1}{d}/{dt}$.
 Table \ref{fNLtable} gives a comparison of both methods of computing
$n_s$. Our results indicate that a significantly larger power
spectrum, along with a generically red-tilted spectrum is an expected
result of curved trajectories in roulette inflation. This is of
particular interest, given that the most recent cosmological data favor a
scalar spectral index of $n_s = 0.96$ \cite{wmap5}.

\subsection{Nongaussianities in the roulette model}
\label{subroulette}

We now turn to the subject of nongaussianity in the roulette model.
Because it gives examples of highly curved field trajectories, one
might have hoped to find observably large levels of nongaussianity
coming from the isocurvature modes.  However, the numerics do
not bear out this expectation, as we now describe.

We tested the algorithm for computing $f_{NL}$ from superhorizon
evolution of perturbation modes on the two-field quadratic inflation
model considered in ref.\ \cite{quantitative}, verified its results.
We then analyzed the nonlinear mode evolution for a variety of
roulette inflation trajectories, for modes $k$ corresponding to a
range of horizon exit times $t_* = \ln{k/H_*}$ before the end of
inflation. 

\begin{figure}
\begin{center}
	\includegraphics[width = .9\textwidth]{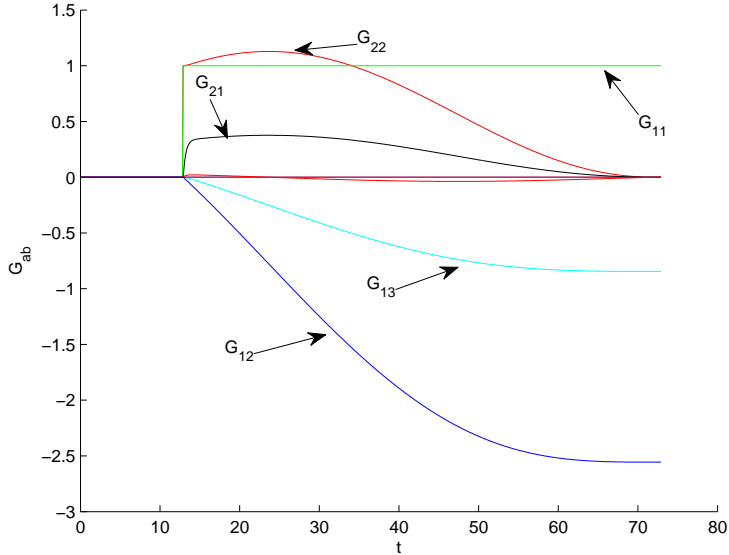}
\end{center}
\vskip-0.9cm
\caption[Green's functions for roulette inflation]{Green's functions $G_{ab}(t,t')$ for $t' = 60$ e-folds before the end of inflation, for roulette inflation with the parameter set 1 and trajectory C. Only the nonzero components are labeled.} 
\label{roulettegreens}
\end{figure}

The Green's function in eq. \eqref{greens2} was found by solving the
ODE numerically in matrix form, as a function of $t$. This was done
once per time step $t'$, giving a $3 \times 3 \times M \times M$
dimensional array, where $M$ corresponds to the number of discrete
time steps sampled in the evolution (typically around
1000).  Figure \ref{roulettegreens} shows this
behavior for trajectory C of figure 2.

Inflation in the $\tau$ valley, corresponding to the effective
one-field scenario of Conlon and Quevedo \cite{kahler} produced no
nongaussianities originating from superhorizon interaction between
scalar modes ($f_{NL} \sim 10^{-20}$, where the deviation from zero
can be attributed to numerical uncertainty).  This is not surprising,
since it is the coupling between curvature and isocurvature modes
that is expected to generate large bispectra.\footnote{Recall that
contributions to $\fNL$ from the adiabatic fluctuations should give
values of order $(n_s -1)$ \cite{maldacena}.} In more complex
inflationary trajectories with sufficient curving in field space,
however, we found that values of $f_{NL}$ between $-0.01$ and $0.02$
were quite generically produced.  But we did not find any examples
which produced $f_{NL}$ outside of the small range
\eqn{
	|f_{NL}| \lesssim 0.1
}
This is far below the level of sensitivity foreseen by the  PLANCK
experiment, for example.

\begin{figure}[ht]
\begin{center}
	\includegraphics[width = .75\textwidth]{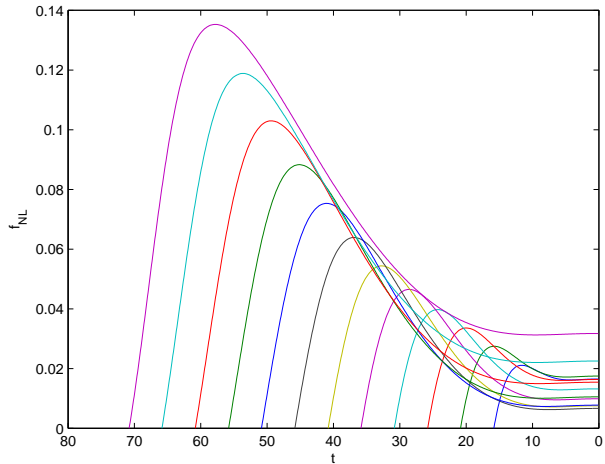}
\end{center}
\caption[$f_{NL}(t)$ for roulette inflation]{Time evolution of the non-linearity parameter $f_{NL}$ for various modes, for trajectory C in the equilateral configuration. The beginning of each curve corresponds to $t = t_*$, the time of horizon crossing of the mode. Here, the x-axis is time until the end of inflation, in e-folds.} 
\label{fNLtdep}
\end{figure}

\begin{figure}
\begin{center}
	\includegraphics[width = .95\textwidth]{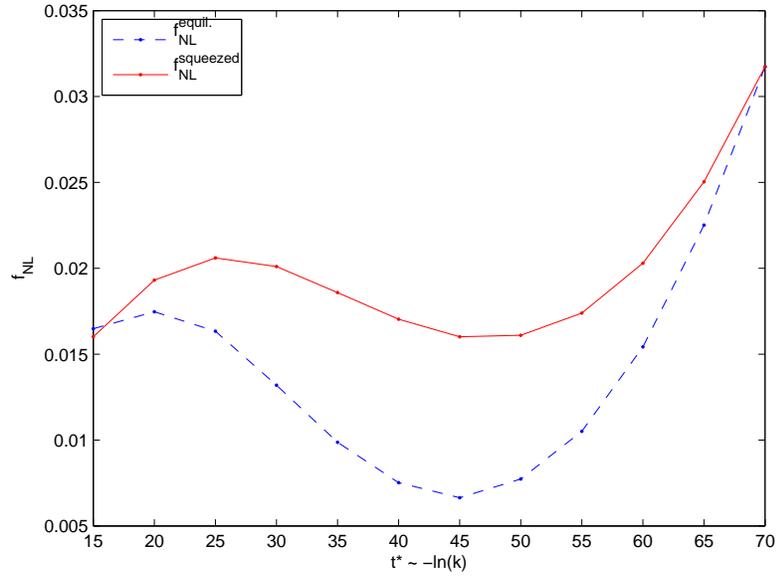}
\end{center}
\caption[$f_{NL}(k)$ for roulette inflation]{Wave-number dependence of
$f_{NL}$ for trajectory C. The x-axis is the time of horizon crossing
in e-folds of the corresponding mode, where $t_* = 0$ is the end of
inflation. In the case of $\fNL^{\rm sqzd.}$ we took $t_{1*}=70$
e-foldings. This is the reason that $\fNL^{\rm equil.}(t_* = 70) = \fNL^{\rm sqzd.}(t_* = 70)$.} 
\label{fNLkdep}
\end{figure}

Figure \ref{fNLtdep} illustrates the time-dependence of $f_{NL}$ at a
series of wave-numbers $k$ from horizon exit to the end of inflation,
for the representative trajectory  C in Figure \ref{trajectories}),
while figure \ref{fNLkdep} shows the $k$-dependence of $f_{NL}$ after
it has stopped evolving. The behaviour of $f_{NL}$ shown is typical
for trajectories that continued to curve during most of 
the period of observable
inflation: they produce slightly
more pronounced values of $f_{NL}$ during the curved part of the
motion, but these values
quickly descend to $\sim 10^{-2}$ by the end of inflation. The
two-field model studied in \cite{quantitative} also exhibits this
behaviour: $\fNL$ descends to zero as the trajectory straightens out
at the end of inflation, resulting in a bispectrum below the level of
measurable sensitivity. Table
\ref{fNLtable} gives some computed values of 
$f^{\rm equil.}_{NL}$ (where the $\vec k_i$ form an equilateral triangle) at
scales $k$ that crossed the horizon at $t_*(k) = 55$ e-foldings,
as well as $f^{\rm sqzd.}_{NL}$ for squeezed triangles, with 
$k_1$, $k_2$ and $k_3$ corresponding to
$t_{*}(k_1) = 60$ and $t_{*}(k_2) = t_{*}(k_3) = 55$ respectively.
Trajectories labeled A through G are shown in figure
\ref{trajectories}. The bottom part of table \ref{fNLtable} shows
results for different parameter sets (given in Table
\ref{paramset}).  Values of $f_{NL}$ are taken at the end of
inflation.

Features in $f_{NL}(k)$ can be understood as being due to curvature
of the trajectory at the time of horizon exit of particular modes.
Modes that experienced more curving after horizon exit
(\textit{i.e.,} those which exited the horizon earlier) produced
larger magnitude $f_{NL}$ than those which experienced no curving.
These features in the bispectrum are completely absent in 
single-field inflation. One feature that was common to curved
roulette trajectories was a slightly larger bispectrum on large
scales, due to the modes which left the horizon before turning in
field space occurred.  This occurs because only superhorizon-scale
modes can experience growth due to the coupling between isocurvature
and adiabatic modes (this effect can also be seen in Table
\ref{fNLtable}, which shows the correlation between large isocurvature
contributions to the power and larger values of $f_{NL}$).
 Smaller scales which exit the horizon later
undergo less such growth.  Thus  a simultaneous detection of larger
$f_{NL}$ in the CMB and smaller primordial non-linearity in
large-scale structure may be a way to detect this type of result, if
its magnitude can be enhanced to an observable level.

One somewhat nonstandard feature of the roulette model, relative to
simpler two-field models, is the  nontrivial field metric $K_{2\bar
2}$ (eq.\ (\ref{k22}) which multiplies the kinetic term  of the
axionic direction $\theta$.  We found that it had no substantial
effect on the shape, size, or magnitude of the bispectrum. To test
this, we tried replacing $K_{2\bar 2}$ by a constant during
inflation, and found that the spectrum of observables remained
roughly unchanged.

As illustrated in Table \ref{fNLtable}, the shape-dependence of the
observed nongaussianities is as expected from this type of model
\cite{shape}, since nongaussianities from superhorizon evolution are
larger in the squeezed configuration, whereas those generated on
subhorizon scales (like in DBI inflation) are larger in the
equilateral configuration.  In all cases, we find that  $f_{NL}$  is
larger for the squeezed configuration.

\section{Discussion}

We have studied a model of \kahler moduli inflation built from a
realistic construction of Type IIB string theory, using the formalism
developed in \cite{nonlin, largenongauss,quantitative}.  We confirm 
previous results for the power spectrum and the superhorizon
influence of isocurvature modes for typical inflationary
trajectories. The main new undertaking is a search for nongaussian
perturbations from highly curved trajectories.  Such deviations from
gaussianity can originate from the superhorizon evolution of the
second-order curvature perturbation and its interaction with
isocurvature perturbations.

When the full spectrum is considered, roulette inflation predicts a
smooth power spectrum with a slight red-tilt, in excellent agreement
with estimates based on the latest WMAP data. In addition, as shown
by \cite{lalak}, the superhorizon influence of isocurvature modes can
come to dominate the scalar curvature power spectrum via the relation
\eqref{gradcons}. This is the case for inflationary trajectories with
large curvature, which can be pictured as a ``projection'' of the
isocurvature perturbation modes onto the adiabatic direction. This
contribution, which can account for over 90\% of the curvature power
spectrum, must therefore be considered in the context of these
multiple-field inflationary models. 

Concerning the issue of nongaussianity, we did not find examples in
which $f_{NL}$, produced by the mechanism considered in section
\ref{sec:results}, could be large enough to  be observable by future
missions; we obtained results which did not exceed the level of
$f_{NL}$ expected in conventional single-field inflation models.  It
was hoped that trajectories with  very sharp curving in field space
would have yielded larger $f_{NL}$ values, but two considerations
could make this difficult. First, if the trajectory relaxes to a
straight, effectively single-field form before the end of inflation, 
$\fNL$ damps to zero,  as illustrated in figure 1 of ref.
\cite{quantitative}. We found trajectories with moderate curving
until the end of inflation, which prevented the total erasure of
$f_{NL}$.   Multiple-field (``multi-brid'') hybrid inflation models
such as discussed in refs.\ \cite{multibrid,Naruko:2008sq} use this
effect to produce sizable nongaussianities. Second, large deviation
from Gaussianity from this method is correlated with large
isocurvature modes being projected onto the curvature direction. If
one is able to find models with large $\fNL$, it must be verified
that the power-spectrum is not overly amplified and distorted by this
effect.  This phenomenon is also seen to play an important role in
nongaussianity generated by the waterfall fields in hybrid inflation
models \cite{hybrid}.

However, it is possible that a more complete search of the model's
parameter space would reveal examples with larger values of
$\fNL$. In that case, running bispectra ({\it i.e.}, which depend
on the magnitude of the average scale of $k_i$, 
as opposed to the shape-dependence), such
as we find in the roulette model, could provide an interesting  
discriminator between models.  Recent developments in the detection of
nongaussianity via large-scale structure, which would probe $f_{NL}$
at smaller scales,  promise to give additional
observational handles on such a dependence \cite{LoVerde, LSS}.

It may be interesting to examine other such moduli inflation
scenarios that arise once the assumptions of a strict hierarchy of
scale are relaxed (for instance, the model of ref.\ \cite{vijay} in
which the second dynamical field is the inverse overall volume,
rather than the axionic partner). A more general study of two-field
inflationary models with such an exponentially flat potential would
furthermore reveal how generic the above-mentioned behavior is.

\section*{Acknowledgments}  We thank N.\ Barnaby, R.\ Brandenberger, F. Cyr-Racine, S.\ Prokushkin, G.\
Rigopoulos and L. Verde  for helpful discussions and correspondence. Our research is supported by NSERC (Canada) and FQRNT (Qu\'ebec).

\appendix

\section{Second order quantities in the gradient expansion formalism}
\label{sec:appendixA}
\numberwithin{equation}{section}
%   \renewcommand{\eqn}{A-\arabic{equation}}
  % redefine the command that creates the equation no.
  \setcounter{equation}{0}  % reset counter 
%   \section*{APPENDIX}  % use *-form to suppress numbering
We present here the important results of \cite{quantitative}, which were used in the computation of the first- and second-order quantities that formed our main results.

Regardless of the specific inflationary model, we may write the
action for a general multifield potential $V(\phi_A)$ 
($A = 1, 2, ...$) with minimal gravitational couplings 
as
\eqn{
	S = \int d^4x \sqrt{-g}\left(\frac{\mplanck^2}{2} R - 
\frac{1}{2}G_{AB}\partial_{\mu}\phi^A\partial^{\mu}\phi^{B} - 
V(\phi^A)\right).
}
Here $R$ is the Ricci scalar, $G_{AB}$ is the metric in field space
and $\mplanck = 1/\sqrt{8\pi G}$ is the reduced Planck mass. We 
work in units where $\mplanck = 1$. At the homogeneous
level, and before specifying a choice of spacelike slicing, the FRW metric is $ds^2 = -N^2(t)\,dt^2 + a^2(t)d\vec{x}^{\,2}$,
and varying with respect to the fields $\phi_A$, $a(t)$ and
$N(t)$ gives the equations of motion for the scalar fields
and the Friedmann equations. Here $N(t)$ is the time-lapse
function, and $a(t)$ is the scale factor, from which we define the
Hubble parameter $H(t) \equiv \dot{a}/(Na)$.  To simplify
calculations during inflation, we make the coordinate choice 
\eqn{
	t \equiv \ln(a),
}
so $N(t) = H^{-1} \simeq$ constant during inflation. Dotted fields will from now on represent differentiation with respect to this time parameter.

Working within the long-wavelength approximation \cite{nonlin}, we
assume that the fields are homogeneous and isotropic within the
horizon. The kinetic term becomes:
\eqn{
	\lagrange_{\rm kin} = \frac{1}{2}G_{AB}\Pi^A\Pi^B,
\label{Lkin}
}
having defined the velocities
\eqn{
	\Pi^A = \frac{\dot{\phi}^A}{N} = H \dot{\phi}^A
\label{momentum}
}

We will furthermore define covariant differentiation of a field that
transforms as a vector within field space \cite{nonlin}:
\begin{eqnarray}
	D_BL^A &=& \partial_BL^A + \Gamma^A_{BC}L^C\\
	D_BL_A &=& \partial_BL_A - \Gamma^C_{AB}L_C\\
	D_{\mu}L^A &=& \partial_{\mu}L^A + \Gamma^A_{BC}\partial_{\mu}\phi^BL^C
\end{eqnarray}
where $\Gamma^A_{BC}$ is the connection defined through the metric
$G_{AB}$. Henceforth, uppercase latin indices $A,B,C, ...$ will
represent the various fields, greek will represent spacetime
indices, and $i,j,k$ will be spatial indices. Since we are interested in
a two-field inflation model, $A,B,C,... = 1, 2$. 
In the roulette model there is only one
independent connection coefficient:
\begin{eqnarray}
	\Gamma^{\tau}_{\tau\tau} = \Gamma^{\theta}_{\tau\theta} = \Gamma^{\tau}_{\theta\tau} = -\Gamma^{\tau}_{\theta\theta} &=&\frac{6\alpha \lambda \tau^{3/2} -\cyvol}{4\tau (\cyvol + 3 \alpha\lambda \tau^{3/2})}\\
 	\Gamma^{\theta}_{\theta\theta} = \Gamma^{\theta}_{\tau\tau} &=& 0 
\end{eqnarray}	

For the analysis of the power spectrum and bispectrum, it will
furthermore be useful to define the orthonormal basis $e_m^A$, where
$m = 1,2$:
\begin{eqnarray}
	e_1^A &=& \frac{\Pi^A}{\Pi},\qquad
	e_2^A = \epsilon_{AB}e_1^B
\label{basis}
\end{eqnarray}
where $\Pi \equiv \sqrt{\Pi_A\Pi^A}$ and $\epsilon_{AB}$ is the
antisymmetric tensor. $e_1^A$ is tangent to the classical field
trajectory,  whereas $e_2^A$ is orthogonal. Note that lower and raised indices $m,n$ are equivalent. 

The scalar field equations of motion are given in eqs.\
(\ref{eom1}-\ref{eom3}).  We numerically integrated them
 to determine the
inflationary  trajectories. The formalism of \cite{nonlin,largenongauss,quantitative} which we follow makes
extensive use of the ``slow roll'' parameters,
\begin{eqnarray}
	\epsilon &=& \frac{\Pi^2}{2H^2} = \frac{\lagrange_{\rm kin}}{H^2} \nonumber\\
	\eta^A &=& -\frac{3H\Pi^A + G^{AB}\partial_BV}{H\Pi} \nonumber\\ 	
	\eta^{\parallel} &=& - 3 - \frac{\Pi^A\partial_AV}{H\Pi^2} \nonumber\\	
	\etaperp &=&  -\frac{e^A_2V_{,A}}{H\Pi}\nonumber \\ 
	\chi &=& \frac{V_{22}}{3H^2} + \epsilon + \eta^{\parallel}\nonumber\\
	\xi_m &=& -\frac{V_{m1}}{H^2} + 3(\epsilon - \eta^{\parallel})\delta_{m1} - 3\eta^{\perp}\delta_{m2}\nonumber\\
	\xipar &=&\xi_1,\quad \xiperp  =\xi_2
\label{params}
\end{eqnarray}
where $V_{22}$ and $V_{m1}$ are defined in the orthonormal basis
(\ref{basis}), such that $V_{mn} = e^A_me^B_nV_{;AB}$, with the covariant derivatives over the field metric defined above. These quantities are nonlinear, depend on both $t$ and $\vec x$, and are not
assumed to be small, although they are small in the slow-roll regime.
It should be noted that some of them are unintuitively named; for
example $\eta^\perp$ is proportional to the slope of the potential, in the
direction orthogonal to the trajectory, rather than a curvature,
and the relation of $\eta^\parallel$ to the usual slow-roll parameters,
with respect to the adiabatic direction, is $\eta^\parallel = -\eta +
\epsilon$.  Nevertheless we will keep this notation for ease of
comparison with ref.\  \cite{largenongauss}.  The $\epsilon$ parameter
does agree with the conventional $\epsilon$ (defined with respect to
the slope along the adiabatic direction).  

The particular choice of variables we will use to describe the fluctuations in this formalism are 
\eqn{
\zeta^A_i(t,\textbf{x}) = e^A_1(t,\textbf{x})\partial_i \ln a(t,\textbf{x}) - \frac{1}{\sqrt{2 \epsilon(t,\textbf{x})}} \partial_i \phi^A(t,\textbf{x})
}
which have the property of being invariant under long-wavelength changes of time-slicing $(t,x) \rightarrow (\tilde t,\tilde x)$ \cite{nonlin}.This quantity can be projected onto the field basis \eqref{basis}:
\eqn{
\zeta_i^m(t,\textbf{x}) = \delta_{m1}\,\partial_i \ln{a} - 
\frac{1}{\sqrt{2\epsilon}}e_{mA}\,\partial_i\phi^A.
}
These simplify in the gauge $t=\ln a$, where $\partial_i \ln{a}=0$. At first order, $\zeta_i^1$ is the spatial gradient of
the usual curvature perturbation, whereas $\zeta_i^2$ corresponds to
the isocurvature perturbation. The gradients are combined, along with their respective velocities $\theta^m_i \equiv
\partial_t\zeta^m_i$, into a 3-component vector,
\eqn{
	v_{ia} = (\zeta^1_i,\zeta^2_i,\theta^2_i)^T,
}
The would-be fourth component is not independent, but is determined
to be 
\eqn{
\theta^1_i = 2\eta^{\perp}\zeta^2_i
\label{gradcons}
}
by the constraint equations  \cite{quantitative}
which may be derived from the Einstein equations and the definition of $\zeta^m_i$, noting that $D_t(\partial_i \phi^A) = D_i(N\Pi^A)$:
\begin{eqnarray}
\partial_i\ln H &=&\epsilon\zeta^1_i, \\
e_{mA} \partial_i \phi^A &= &-\sqrt{2\epsilon} \zeta^m_i, \nonumber \\
e^A_m D_i\Pi_A & = &-H\sqrt{2\epsilon}\left(\theta^m_i + \etapar\zeta^m_i - \etaperp \zeta^2_i\delta_{m1} +(\etaperp\zeta^1_i +\epsilon\zeta^2_i)\delta_{m2}\right). \nonumber
\end{eqnarray}
The relationship between $\theta^1_i$ and $\zeta^2_i$ is nothing more than the well-known ``conservation'' law of the curvature perturbation. This is valid to all orders, as shown in ref.\ \cite{conservation}.

Combining these with the equations of motion
\eqref{eom2}-\eqref{eom3}, the full nonlinear evolution equations may
be written in the compact form:
\eqn{
\dot{v}_{ia}(t,\textbf{x}) + A_{ab}(t,\textbf{x})v_{ib}(t,\textbf{x}) = 0.
\label{nonlinear}
} 
The matrix $A$ is a function of the parameters defined in eq.\
\eqref{params} \cite{quantitative}:
\eqn{
	A = \left(\begin{array}{ccc}
		0 & -2\eta^{\perp} & 0\\
		0 & 0 & -1\\
		0 & 3\chi + 2\epsilon^2 +4\epsilon\eta^{\parallel} 
+ 4(\eta^{\perp})^2 + \xi^{\parallel} -2\epsilon R_{2112} & 3 
+ \epsilon + 2\eta^{\parallel}
\end{array} \right)
\label{Aeq}
}
Its dominant components are $A_{33}\cong 3$ and $A_{23}=-1$. The only
explicit dependence on the curvature of the field manifold in $A$ is
the term $ -2\epsilon R_{2112} \equiv -2\epsilon
e^A_2e^B_1e^C_1e^D_2R_{ABCD}$, but we found that this is negligible
 ($\sim 10^{-6}$) in roulette inflation.

The next step is to solve this system of equations perturbatively. Eq.\ \eqref{nonlinear} can be expanded into a hierarchy of linear perturbation equations for $v_{ia}^{(n)}$, each sourced by the previous order. 
Since we are interested in superhorizon evolution, it
is reasonable to take the first-order perturbations to be sourced 
by a linear perturbation $b^{(1)}_{ia}$, which
encodes the effect of quantum fluctuations at short wavelengths
providing the initial values for the long-wavelength modes of 
interest at horizon crossing.  Refs.\ \cite{largenongauss,nonlin,quantitative} show that the source term
having the right properties is
\eqn{
	b^{(1)}_{ia} = \int \frac{\ud^3\textbf{k}}{(2\pi)^{2/3}} \ 
\dot{\window}(k)X^{(1)}_{am}
\hat a^{\dagger}_m(\textbf{k})i k_i\e{i \mathbf{k\cdot x}} + 
\mathrm{c.c.},
}
where the creation operator has the standard commutator
$[\hat a_m(\textbf{k}),\hat a^{\dagger}_n(\textbf{k}')] = 
\delta_{mn}\delta^{(3)}(\textbf{k}-\textbf{k}').$
Superscripts in parentheses indicate the expansion order in
perturbation theory.

The matrix of linear solutions around horizon
crossing $X_{am}$ is the slow-roll solution of ref.\
\cite{quantitative} in which it is argued that deviations from
linearity on sub-horizon scales should be slow-roll suppressed:
\eqn{
	X_{am} = -\frac{H}{4k^{3/2}\sqrt{\epsilon}}\left(\begin{array}{cc}
	                                            1 & 0\\
					            0 & 1\\
						 0 &-\chi
	                                           \end{array}\right).
\label{Xeq}
}
The factor $1/\sqrt{2\epsilon}$ comes from the definition of $\zeta$, and the amplitude $H$ is the result we expect from perturbations at horizon crossing. 

The window function $\window(t,k)$ is designed to source only the
superhorizon modes, and the final
results must be independent of its exact shape.  It is convenient to
use a Heaviside step function, $\window(t,k)=\Theta(kR-1)$, that
has support only on scales $R = (c/aH) = (c/H)\e{-t}$ (recall we are
in the gauge $t = \ln a$)
sufficiently larger than the Hubble radius, where $c$ should be of
order a few. Given that fluctuations that are generated on sub-horizon scales do not yet feel the effect of curvature, and therefore correspond to fluctuations in Minkowski space, it is reasonable to expect the spectrum of fluctuations on these scales to be Gaussian. 

 Then
\eqn{
	\dot\window(t,k) = \delta(kR  -1) = 
	\frac{\delta(t - t_* - \ln c)}{\vert- c\e{-t+t_*}\vert},
}
where $t_*$ is the time of horizon-crossing of mode $k$
\eqn{
	t_* \equiv \ln{k/H_*}.
}
Physical quantities are found to be independent of the exact value 
of $c>1$.

The first- and second-order equations can then be written:
\begin{eqnarray}
	\dot{v}_{ia}^{(1)}(t,\textbf{x}) + A_{ab}^{(0)}(t,\textbf{x})v_{ib}^{(1)}(t,\textbf{x}) &=& b^{(1)}_{ia},\label{firstorder} \\
	\dot{v}_{ia}^{(2)}(t,\textbf{x}) + A_{ab}^{(0)}(t,\textbf{x})v_{ib}^{(2)}(t,\textbf{x}) &=&  -A_{ab}^{(1)}(t,\textbf{x})v_{ib}^{(1)}(t,\textbf{x}). 
\end{eqnarray}
Here, $A_{ab}^{(1)} = \bar{A}_{abc}^{(0)}(t)\partial^{-2}\partial^i
v_{ic}^{(1)}$ (note that
$\partial^{-2}$ is just multiplication by $-k^{-2}$ in Fourier
space.) $\bar{A}$ is given by \cite{quantitative}:
\eqn{
	\left(\!\!\!\begin{array}{ccc}
		\mathbf{0} & \left(\!\!\!\!\!\begin{array}{c} 2\epsilon\etaperp -4\etapar\etaperp + 2\xiperp \\ -6\chi -2\epsilon\etapar - 2(\etapar)^2 - 2(\etaperp)^2 \\ -6 - 2\etapar \end{array} \!\!\!\!\!\right) & \mathbf{0} \\

		\mathbf{0}  & \mathbf{0}  & \mathbf{0} \\
		\mathbf{0}  & \mathbf{\bar{A}_{32}} &\!\!\!\!\!\!\!\!\!\!\left(\!\!\!\!\begin{array}{c} -2\epsilon^2 - 4\epsilon\etapar + 2(\etapar)^2 - 2(\etaperp)^2 -2\xipar \\ -4\epsilon\etaperp -2\xiperp \\ -2\etaperp \end{array} \!\!\!\right)
\end{array} \!\!\!\!\!\right)
\label{barA}
}
where
\eqn{
\mathbf{\bar{A}_{32}} \equiv- 2\partial_i \epsilon R_{2112} + \left(\begin{array}{c} \mathbf{\bar{A}_{321}} \\
							\mathbf{\bar{A}_{322}} \\
							\mathbf{\bar{A}_{323}} 
\end{array} \right)
}
and
\begin{eqnarray}
\mathbf{\bar{A}_{321}} &\equiv& -6\epsilon\etapar -6(\etaperp)^2 -3\epsilon\chi \\
 &&-4\epsilon^3 -10\epsilon^2\etapar -2\epsilon(\etapar)^2 \nonumber\\
&&-6\epsilon(\etaperp)^2 + 8\etapar(\etaperp)^2 - 3\epsilon\xipar \nonumber \\
 &&- 6\etaperp\xiperp + \sqrt{\frac{\epsilon}{2}}(V_{111} - V_{221}) \nonumber \\
\mathbf{\bar{A}_{322}} &\equiv& -12\epsilon\etaperp -6\etapar\etaperp  \\
 &&+ 12\etaperp\chi -6\epsilon^2\etaperp + 4(\etaperp)^3  \nonumber \\
&&-4\epsilon\xiperp -2\etapar\xiperp + \sqrt{\frac{\epsilon}{2}}(V_{211} - V_{222}) \nonumber \\
\mathbf{\bar{A}_{323}} &\equiv& 6\etaperp - 2\epsilon\etaperp + 4\etapar\etaperp -2\xiperp 
\end{eqnarray}
The first index in $\bar A_{abc}$ stands for the row, the second for
the column in (\ref{barA}), and the third for the ``depth'' dimension
of the array, represented here by a column vector for each $\bar A_{ab}$.
$V_{lmn}$ is defined as $V_{lmn} \equiv e^A_l e^B_m e^C_n V_{;ABC}$. 

The above linear equations can then be solved with the aid of the Green's
function which is the solution to the inhomogeneous equation
\eqn{
\frac{\ud}{\ud t}G_{ab}(t,t') + A_{ac}^{(0)}(t)G_{cb}(t,t') = \delta(t-t').
\label{greens}
}
with $G_{ab}(t,t) = \delta_{ab}$ at equal times.
This must be solved only once for each classical trajectory, which we
do numerically on a grid in $t$, $t'$, $a$ and $b$. 
Once $G_{ab}(t,t')$ is known, the step-function form of $\window(t,k)$
simplifies the integration of the first order solution,
\begin{eqnarray}
v_{am}^{(1)}(k,t) &=& \int_{-\infty}^t \ud t'\,G_{ab}(t,t')\,
\dot \window(k,t')\,X^{(1)}_{bm}(k,t')\nonumber\\
		&=& G_{ab}(t,t_* + \ln c)\,X^{(1)}_{bm}(k,t_* + \ln c), 
\label{veq}
\end{eqnarray}
% 
% where
% \eqn{
% 	v_{ia}^{(1)}(\mathbf x,t) = \int \frac{\ud d^3 \mathbf k}{(2\pi)^{3/2}}v_{am}^{(1)}(k,t)\hat 
% }

where we have defined the Fourier-space perturbation as:
\begin{eqnarray}	
	v_{ia}^{(1)}(\mathbf x,t) &=& \partial_i v_a^{(1)} = \int \frac{\ud^3\textbf{k}}{(2\pi)^{2/3}}\,
v^{(1)}_{am}(k,t) \, a^{\dagger}_m(\textbf{k})\,i k_i\,
\e{i \mathbf{k\cdot x}} + \mathrm{c.c.}
\end{eqnarray}

The second order solution can be expressed, using the same method, as:
\eqn{
v^{(2)}_{ia}(\mathbf x,t) = -\int \ud t'\, G_{ab}(t,t')\,
\bar A_{bcd}(t')\, v_{ic}^{(1)}(\mathbf x,t')\,
\partial^{-2}\partial^j v_{jd}^{(1)}(\mathbf x,t').
\label{secondorderintegral}
}
To connect with observables one transforms the time coordinate $t$ in
the gauge of uniform expansion time slices ($NH = 1$) to  $T(t,x)$ which
describes uniform density slices ($\partial_i \rho = 0$)
\cite{quantitative},  so $a(t)\to \tilde a(T,x)$.  Then the curvature
perturbation can be expressed as the total gradient of a scalar $\tilde \alpha$, 
$\tilde\zeta^1_i  = \partial_i \ln \tilde a \equiv 
\partial_i\tilde\alpha$, which allows observable scalar correlators
to be expressed simply. Note that this result should be identical to results found using the $\delta N$ formalism, given that the perturbation $\delta \tilde \alpha$ in the uniform density gauge corresponds exactly to the perturbation in the number of e-folds $\delta N \equiv \delta \ln \tilde a = \zeta$ \cite{Wands:2000dp}.

The curvature power spectrum is:
\eqn{
\mathcal{P}(k,t) = \frac{k^3}{2\pi^2}\langle \tilde \alpha 
\tilde \alpha \rangle(k,t) = 
\frac{k^3}{2\pi^2}{v}_{1m}^{(1)}(k,t){v}_{1m}^{(1)}(k,t).
\label{fullP}
}
The scale-dependence of $P$ comes as expected from the time-dependence
of $H$ in $X_{am}(k,t)$ (eq.\ (\ref{Xeq})), which appears in 
${v}_{1m}^{(1)}(k,t)$ through eq.\ (\ref{veq}). It should be stressed that the power spectrum here is complete and includes the effect of isocurvature perturbations.

The leading 
contribution to the  bispectrum comes from the expansion to second
order in perturbation theory,
\begin{eqnarray}
\label{bispectrum}
\langle \tilde{\alpha}_{k_1} \tilde{\alpha}_{k_2} \tilde{\alpha}_{k_3}
\rangle^{(2)}(t) &=& \langle \tilde{\alpha}_{k_1}^{(1)}
\tilde{\alpha}_{k_2}^{(1)} \tilde{\alpha}_{k_3}^{(2)}
\rangle(t) +  (k_1 \leftrightarrow k_3) 
+  (k_2 \leftrightarrow k_3)\\
&=& (2\pi)^3\delta^3(\mathbf{k_1} + \mathbf{k_2} + 
\mathbf{k_3})\left[ f(k_1,k_2) + f(k_2,k_3) + f(k_1,k_3) \right] \nonumber
\end{eqnarray}
where \cite{quantitative}
\eqn{
f(k,k') \equiv \left(\frac{1}{2}v^{(2)}_{1mn}(k,k',t) +  
\etaperp v_{2m}^{(1)}(k,t)v_{1n}^{(1)}(k',t)\right)
v_{1m}^{(1)}(k,t)v_{1n}^{(1)}(k',t) + k \leftrightarrow k'.
\label{fkkp}
}
The second term in parentheses comes from the coordinate change
$t\to T$, and the first term is given by
\eqn{
	v^{(2)}_{1mn}(k,k',t) \equiv -\int_{-\infty}^{t} \ud t'\,
 G_{1a}(t,t')\,\bar A_{abc}(t')\,v_{bm}^{(1)}(k,t')\,
v_{cn}^{(1)}(k,t').
\label{veq2}
} 
Numerically, we will find that this term dominates over the  $\etaperp
v_{2m}^{(1)}(k,t)v_{1n}^{(1)}(k',t)$ term in the roulette inflation
model by five orders of magnitude. These are all the ingredients 
needed for evaluation of the nonlinearity parameter 
$f_{NL}$ \cite{largenongauss},
\eqn{
f_{NL} = \frac{\langle\alpha^{(1)}_{k_1}
\alpha^{(1)}_{k_2}\alpha^{(2)}_{k_3}\rangle +  (k_1 \leftrightarrow k_3) 
+  (k_2 \leftrightarrow k_3)}
{\langle \alpha^{(1)}\alpha^{(1)}\rangle_{k_1}\langle 
\alpha^{(1)}\alpha^{(1)}\rangle_{k_1} +  (k_1 \leftrightarrow k_3) 
+  (k_2 \leftrightarrow k_3)}. 
}

% \bibliography{nongauss.bib}
% \bibliographystyle{plain}

\end{document}